 \documentclass[prb,superscriptaddress,preprint,showpacs,showkeys]{revtex4}
\usepackage{graphicx}
\usepackage{bm}
\usepackage{amsmath}
\usepackage{amssymb}

\begin{document}

\title{Electrostatic deflections of cantilevered semiconducting single-walled carbon nanotubes}

\author{Z. Wang}
\email{zhao.wang@univ-fcomte.fr}
\affiliation{Institut UTINAM, UMR 6213, University of Franche-Comt\'{e}, 25030 Besan\c {c}on Cedex, France.}
\author{M. Devel}
\affiliation{Institut UTINAM, UMR 6213, University of Franche-Comt\'{e}, 25030 Besan\c {c}on Cedex, France.}
\author{R. Langlet}
\affiliation{Laboratoire de Physique du Solide, FUNDP, Rue de Bruxelles 61, 5000 Namur, Belgium}

\author{B. Dulmet}
\affiliation{FEMTO-ST, UMR 6174, DCEPE, ENSMM,  25030 Besan\c {c}on Cedex, France}

\begin{abstract}
How carbon nanotubes behave in an external electric field? What will be the relation between the intensity of the electric field and the tube's deformation? What are the geometry effects on the response of carbon nanotubes to electric fields? To answer these questions, we have developed a new combined computational technique to study electrostatic field induced deformations of carbon nanotubes. In this work, we find that the deflection angle of cantilevered semiconducting single-walled carbon nanotubes is  proportional to the square of the electric field strength, and the tubes can be most bent when the field angle ranges from $45$ to $60$ degrees. Furthermore, the deflection angle is also found to be proportional to the aspect ratio $L/R$. Our results provide a good qualitative agreement with those of one previous experimental study.
\end{abstract}

\pacs{73.63.Fg, 77.22.Ej, 31.15.-p, 77.55.+f}

\keywords{carbon nanotubes, electrostatic deflection, polarization, dipole, induced deformation, NEMS, dielectric properties}

\date{\today}

\maketitle

\section{Introduction}
\label{intro}

Carbon nanotubes (CNTs) can be ideal building blocks for nanoelectromechanical systems (NEMS) due to their unique electrical and mechanical properties \cite{saito-98-book}. Thus, they have attracted much interest from both technical and scientific communities concerned with sensing, actuation, vibration and lab-on-a-chip applications \cite{ke-05-book}. Cantilevered semiconducting CNTs can be used as key elements in NEMS such as nanotweezers \cite{akita-01} and nanorelays \cite{kinaret-03}. Their electric field induced deformation is a key character for these promising applications, as well as their fabrication \cite{li-04}, separation \cite{krupke-03} and electromanipulation \cite{hertel-98}. 

In a previous study, Poncharal \textit{et al} \cite{Poncharal-99} reported the electric deflections of cantilevered multiwalled CNTs observed using a transmission electron microscope. Electrostrictions in single-walled CNTs (SWCNTs) were observed by El-Hami and Matsushige \cite{el-hami-05} using an atomic force microscopy (AFM). Q.-L. Bao \textit{et al} \cite{bao-01} reported that the microstructure of CNTs can be changed by an electric field during growth using a high-resolution transmission electron microscope. Y.-F. Guo and W.-L. Guo \cite{guo-03} carried out quantum mechanics calculations to investigate the electric field induced tensile breaking and the influence of external field on the tensile stiffness of CNTs, and found that both the tensile stiffness and the strength of CNTs decrease with increasing intensity of electric field. Torrens \cite{torrens-04} reported that the polarizabilities of CNTs can be modified reversibly by external radial deformation. Mayer and Lambin \cite{mayer-05-03} calculated the electrostatic forces acting on CNTs placed in the vicinity of metallic protrusions for dielectrophoresis. W. Guo and Y. Guo\,\cite{guo-03-02} found giant electrostrictive deformations in SWCNTs using quantum mechanics simulations.

The calculations using traditional first-principle methods are very time-consuming, so that the size of the systems considered is always very limited in past theoretical studies concerning the effects of external field. That is one of the main reasons why deformations of CNTs in electric fields and their geometric effects are still not fully understood up to date. 

How CNTs behave in an external electric field? What will be the relation between the intensity of the electric field and the tube's deformation? What are the geometry effects on the response of CNTs to electric fields? To answer these questions, we have developed a new combined computational technique to study electrostatic field induced deformations. In this technique the interatomic potential is described in an empirical way and the electrostatic interaction is calculated using a Gaussian renormalized point dipole interaction model (e.g. interactions between broaden dipoles). The main advantage of this method is its ability to deal with much larger systems with a reasonable computational requirement. This enabled us to carry out a series of simulations to study the induced deformations of various carbon structures in electrostatic fields.    

In this work, we study the electrostatic deflections of cantilevered semiconducting SWCNTs in a non-axial homogeneous field. This study addresses the relation between the external field, the mechanical resistance of SWCNTs and their electrostatic polarizabilities. In the last few years, the polarizability of CNTs has been intensively investigated \cite{benedict-95,mayer-05-01,mayer-05-02,guo-04,kozinsky-06,novikov-06}. For example, it is known that the static polarizabilities of CNTs are very anisotropic, and that the polarizabilities are proportional to the radius square in the plane perpendicular to the tube axis and almost independent of chirality. In this study we use Gaussian regularized propagators \cite{langlet-04}$^,$ \cite{langlet-06} to calculate the local polarizability of carbon atoms. In this model, the standard vacuum propagator is convoluted by a Gaussian function, in order to avoid polarization catastrophes. This is the same as considering that the dipoles are not real point dipoles but rather due to Gaussian distribution of charges whose width is related to the polarizabilities \cite{mayer-06-01}$^,$\cite{mayer-07-01}. Compared to first-principle techniques such as \textit{ab initio} calculations, the local auto-coherent polarizabilities related to the geometrical structure of carbon atoms can be quickly evaluated in this model. It can therefore be practically used in dynamic simulations for larger systems.

In this work, a simple gradient algorithm is used to calculate equilibrium configurations of the atoms by minimizing the total energy of the systems, which consists of both the interatomic and the induced electrostatic interactions. The interatomic potential is analyzed using an adaptive intermolecular reactive empirical bond-order hydrocarbon (AIREBO) potential function formulated by Stuart \textit{et al} \cite{stuart-00}. During the simulations, deflections of cantilevered CNTs are recorded dynamically as a function of the iteration step. Examples can be viewed at one of the authors' personal web page \cite{Wang-site}. The theory will be presented in section 2. In section 3, we examine the relations between the deflections and the intensities and orientations of the applied fields, respectively. The effects of tubes chiralities, radii and lengths are also studied. We draw a conclusion at the end of this section.

\section{Theory}

In this study, we examine cantilevered semiconducting SWCNTs submitted to an external homogeneous electrostatic field in free space with zero net charge and zero permanent dipole moment. At the beginning of the simulations, an open-ended tube, fixed at one end, is relaxed. When an electric field is applied, a dipole is induced on each atom. The induced dipole is modeled as an ideal dipole with anisotropic linear polarizabilities. The field intensity is supposed small enough so that any hyperpolarization effect can be neglected. A simple gradient algorithm is used to simulate the deformation of CNTs in the field by minimizing the total energy of the systems $U^{tot}$. The motionless equilibrium configuration of the atoms corresponds to the minimum value in the energy curve. Classical molecular dynamics is not used in order to avoid the thermal vibration because we consider only the equilibrium stats of tubes in this study. $U^{tot}$ is the sum of two terms: the induced electrostatic energy $U^{elec}$ and the interatomic potential $U^{p}$.   

\begin{equation}
\label{eq:1}
U^{tot}=U^{elec}+U^{p}
\end{equation}

In which $U^{elec}$ consists of three terms:

\begin{equation}
\label{eq:2}
U^{elec}=\frac{1}{2}\sum_{i=1}^N{\bm{p}_i\alpha_i^{-1}\bm{p}_i}
-\frac{1}{2}\sum_{i=1}^N{\sum_{\substack{j=1 \\ j\ne i} }^N{\bm{p}_i\bm{E}_j(\bm{r}_i)}}-\sum_{i=1}^N{\bm{p}_i\bm{E}_0(\bm{r}_i)}
\end{equation}

where $N$ is the total number of atom, $\bm{p}_i$ is the dipole induced on atom $i$,  $\bm{E}_j({\bm r}_i)$ represents the electric field created by an other dipole $\bm{p}_j$ around atom $i$, $\bm{E}_0({\bm r}_i)$ stands for the external field, $\alpha_i$ is the local anisotropic polarizability tensor of atom $i$ adapted from graphite \cite{langlet-06}. In this expression, the first term is the self energy term, the second term accounts for the dipole-dipole interaction, the last term presents the interaction with the external field \cite{mayer-06-01}. The two basic parameters of the polarizability tensor are taken to be:  $\alpha_{//}=2.47$\AA$^3$ and $\alpha_\bot =0.86$\AA$^3$, with $\scriptstyle//$ and $\scriptstyle\bot$ meaning parallel and perpendicular to the plane of graphene sheet, respectively. Note that due to the use of point dipole model, these parameters are not valid for metallic tubes.

The field on atom $i$ due to another induced dipole $\bm{E}_j$  can be written as:

\begin{equation}
\label{eq:3}
\bm{E}_j(\bm{r}_i)=\bm{T}_2(\bm{r}_i, \bm{r}_j)\alpha_j\bm{E}_l(\bm{r}_j)=\bm{T}_2(\bm{r}_i,\bm{r}_j)\bm{p}_j
\end{equation}

In this equation, $\bm{E}_l(\bm{r}_i)$ describes the local field at atom $i$ and the regularized tensor $\bm{T}_2$ is the double gradient of Green's generalized function for the Laplace equation, convoluted by a Gaussian distribution to avoid polarizability catastrophes \cite{langlet-04}$^,$\cite{langlet-06}. The distribution of dipoles {$\bm{p}_i$} is determined by the fact that the equilibrium state of the dipole distribution should correspond to the minimum value of $U^{elec}$. This means that the partial derivatives of the total electrostatic energy with respect to the $3 \times N$ components of the dipoles should be zero. 

\begin{equation}
\label{eq:4}
\forall{i=1,...,N}\,\,\,\,\,\,\,\,\,\,      \frac{\partial{U^{elec}}}{\partial{p_{i,x}}}=\frac{\partial{U^{elec}}}{\partial{p_{i,y}}}=\frac{\partial{U^{elec}}}{\partial{p_{i,z}}}=0         
\end{equation}

Where $p_{i,x}$, $p_{i,y}$, $p_{i,z}$ are the three components of the dipole $\bm{p}_i$. Putting Eq.\,\ref{eq:2} and Eq.\,\ref{eq:3} into Eq.\,\ref{eq:4} gives a linear system of $N$ equations for $N$ vectorial variables. 

\begin{equation}
\label{eq:5}
\bm{p}_i= \alpha_i\bm{E}_0(\bm{r}_i)+\sum_{\substack{j=1 \\ j\ne i} }^N\alpha_i\bm{T}_2(\bm{r}_i, \bm{r}_j)\bm{p}_{j}
\end{equation}

The local fields and the dipole moments can be calculated by solving this system. By putting these expressions for the minimizing $\bm{p}_{i}$ back into Eq.\,\ref{eq:2}, cancellation of various terms leaves only a simple expression for the minimum value of $ U^{elec}$ as a function of the $\bm{p}_{i}$:
\begin{equation}
\label{eq:6}
U^{elec}_{min}=-\frac{1}{2}\sum_{i=1}^N\bm{p}_{i}\bm{E}_0(\bm{r}_i)
\end{equation}

The interatomic potential $U^p$ is determined using the AIREBO potential function \cite{stuart-00}. This potential is an extension of Brenner's second generation potential \cite{brenner-02} and includes long-range atomic interactions and single bond torsional interactions. In this type of potential, the total interatomic potential energy is the sum of individual pair interactions containing a many-body bond order function:

\begin{equation}
\label{eq:7}
U^p=\frac{1}{2}\sum\limits_i {\sum\limits_{j\ne i} 
{\left[ 
\begin{array}{l}
V^R(r_{ij})-b_{ij} V^A(r_{ij} ) + V_{ij}^{L-J}(r_{ij}) 
 + \sum\limits_{k\ne i,j} 
{\sum\limits_{\ell\ne i,j,k} {V_{kij\ell}^{tor} } }
\end{array}
 \right]} }
\end{equation}

where $V^R$ and $V^A$ are the interatomic repulsion and attraction terms between valence electrons, for bound atoms. The bond order function $b_{ij}$ provides the many body effects by depending on the local atomic environment of atoms $i$ and $j$. The long-range interactions are included by adding a parameterized \textit{Lennard-Jones} 12-6 potential term $V^{L-J}$ . $V^{tor}$ represents the torsional interactions.

\section{Results and discussion}

In this study, we use Cartesian coordinates with the $z$ axis along the principal axis of the tube. The tubes are fixed at one end in an imposed homogeneous electric field. The tube length is in the range of 2.0 to 8.4 nm. In order to get enough measurable deflection of these short SWCNTs, the strength of applied external electric fields is in the range of $0.5$ to $3.0$\,V/nm. This is much stronger than the ordinary field reported in the experiments. On the other hand, as going to be discussed in the following parts of this section, the deflection of SWCNTs is proportional to the square of tube length and it is also proportional to the square of field strength. For example, in order to get the same electrostatic deflection for a tube one thousand times longer, it is just enough to apply a field a thousand times smaller. Hence, considering that the lengths of CNTs studied in previous experimental works are in the ranges of some micrometers, the field emission effect caused by strong fields is neglected. Furthermore, we note that the actual external electric field created by the electrodes is usually inhomogeneous in real experiments, however this inhomogeneous distribution varies from one experimental setup to another. For the sake of simplicity, we have therefore chosen to use a homogeneous external field distribution, as in previous theoretical studies.

In this work, all applied electric fields are parallel to the $y-z$ plane due to the symmetry of the system. The field angle $\theta$ is defined as the angle between the field direction and the $z$ axis. The deflection at the tip of the tube is noted as $w$. We studied at first the relation between the intensity of the external field and the deflection of cantilevered SWCNTs. 

\begin{figure}[tb]
\centerline{\includegraphics[width=8cm]{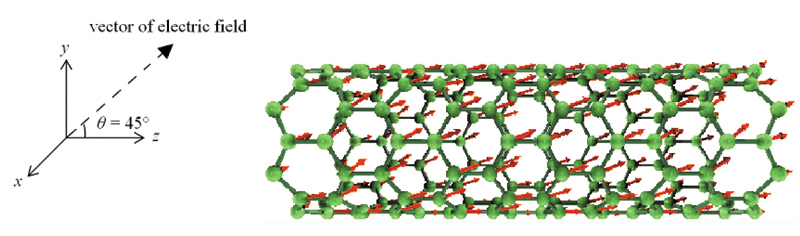}}
\caption{\label{fig:1}
(Color online) Induced dipoles in a (8,0) tube (with non-optimized structure) in an electrostatic field. Field angle $\theta=45^\circ$. Field strength $E=0.71$\;V/nm. Tube length $L=1.988$\;nm. The green points stand for atoms. The arrow with a dashed line on the left presents the electric field vector in the $y-z$ plane. The red arrows with solid lines on the tube stand for the induced dipoles.}
\end{figure}

When a carbon atom is brought into an electric field, the field tends to shift the electrons and the nuclei in opposite directions. Thus, induced dipoles are created. Fig.\,\ref{fig:1} shows dipoles induced on atoms by an electrostatic field in a zigzag tube (8,0). It is selected because of its symmetric geometry and small radius.

\begin{figure}[tb]
\centerline{\includegraphics[width=8cm]{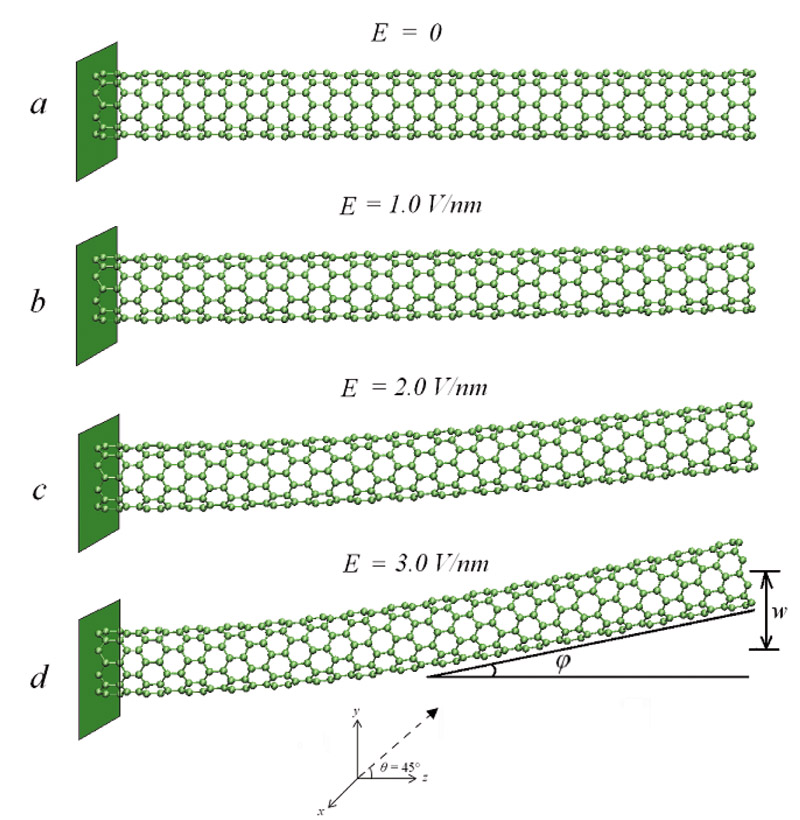}}
\caption{\label{fig:2}
(Color online) Induced deflections of a (8,0) cantilevered tube in electrostatic fields with different field intensities. The direction of the applied electric fields is parallel to the $y-z$ plane with $\theta=45^\circ$. The original length of the tube is about $6.54$\,nm. The arrow with a dashed line represents the electric field vector. Fig.\,(\textit{a}) shows a tube relaxed in vacuum without any electric field. Figs.\,(\textit{b})-(\textit{d}) show the deflected shape of the tubes in various applied external electric fields ($E=1.0-3.0$\,V/nm).}
\end{figure}

The positive and negative charges of CNT are brought about a relative displacement due to the electric force. The tube tends to be either parallel (in most cases) by dipole-dipole and dipole-field interactions. Tubes can therefore be bent by induced forces and moments to the field direction. Fig.\,\ref{fig:2} shows the deflections of a CNT induced by various external electric fields corresponding to various field strength $E$, but a given $\theta=45^\circ$. Note that these are the equilibrium configurations of the atoms, around which the tubes would vibrate at non-zero temperature\,\cite{treacy-96}.

\begin{figure}[tb]
\centerline{\includegraphics[width=8cm]{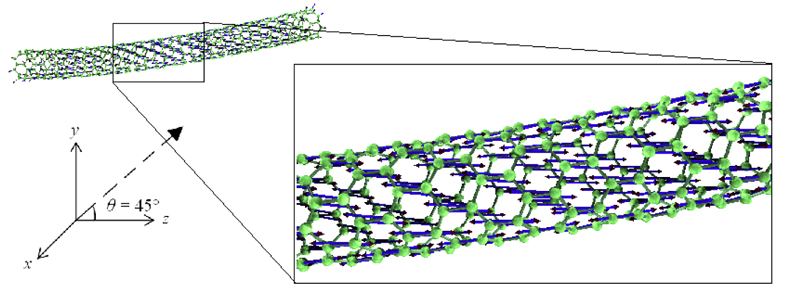}}
\caption{\label{fig:3}
(Color online) Vectors of induced forces in a (8,0) tube bent by an electrostatic field. $\theta=45^\circ$. $E= 4.0$\,V/nm. $L=6.54$\,nm. The green point represents the carbon atoms. The arrow with a dashed line represents the external electric field vector. The blue arrows on the atoms stand for the induced forces, which are calculated as the negative gradients of the electrostatic energy.}
\end{figure}

It can be seen that as expected, the tubes are bent toward the direction of the external electric field. Note that the tube is only curved at the part closed to the fix end. The right side part of the tubes remains straight. We compared the atomic structure of this straight part in Fig.\,\ref{fig:2}\,(\textit{b})-(\textit{d}) to an unmoved tube as shown in Fig.\,\ref{fig:2}\,(\textit{a}). It is found that the average bond length in this part is slightly smaller than that in the original one. That means the tube is not only bent by induced moments but also compressed by induced forces at the same time, as the electrostriction effects found in Refs.\,\onlinecite{el-hami-05} and Refs.\,\onlinecite{guo-03-02}. In fact, the distribution of electrostatic forces acting on the atoms is rather inhomogeneous as shown in Fig.\,\ref{fig:3} as an example. In consequence, the macroscopic continuum mechanics models of beam structures subjected to simple loadings can not be directly used to calculate internal strains for tubes in an electric field, without incorporating volume densities of torques. 

\begin{figure}[tb]
\centerline{\includegraphics[width=8cm]{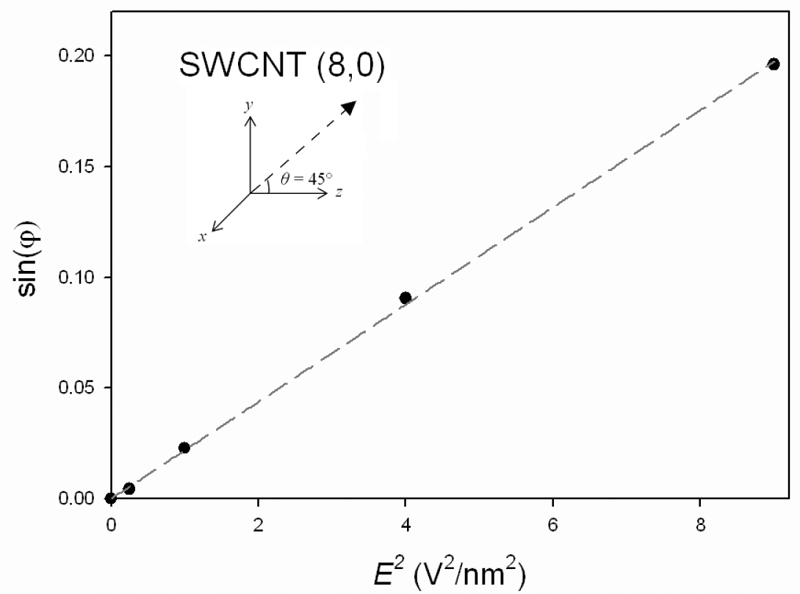}}
\caption{\label{fig:4}
(Color online) Sine of deformation angles $\sin(\varphi)$ versus the square of the electric field strength $E$ for a (8,0) tube. The tube length is about $6.54$\,nm. $\theta=45^\circ$. The round points stand for the simulation data and the green dashed line represents the fitting line.}
\end{figure}

For convenience, we define a deformation angle $\varphi$ to describe the deformations of the tubes. It is equal to the angle between the $z$ axis and the neutral axis of the deformed CNTs at their free end. The relation between $\varphi$ and the intensity of the external field is presented in Fig.\,\ref{fig:4}.

It can be seen that $\sin(\varphi)$ is almost proportional to the square of $E$. 

\begin{equation}
\label{eq:8}
\sin(\varphi)=A E^2
\end{equation}

Where $A = 0.0219 \pm 0.0002$ for this (8,0) tube with $\theta=45^\circ$. Note when the value of $\varphi$ remains small, one can also find that the deflection at the tip  of a long tube $w=L \times \sin(\varphi)$ is almost proportional to $E^2$. This provides a good qualitative agreement with the experimental study of Poncharal \textit{et al} \cite{Poncharal-99}.

\begin{figure}[tb]
\centerline{\includegraphics[width=8cm]{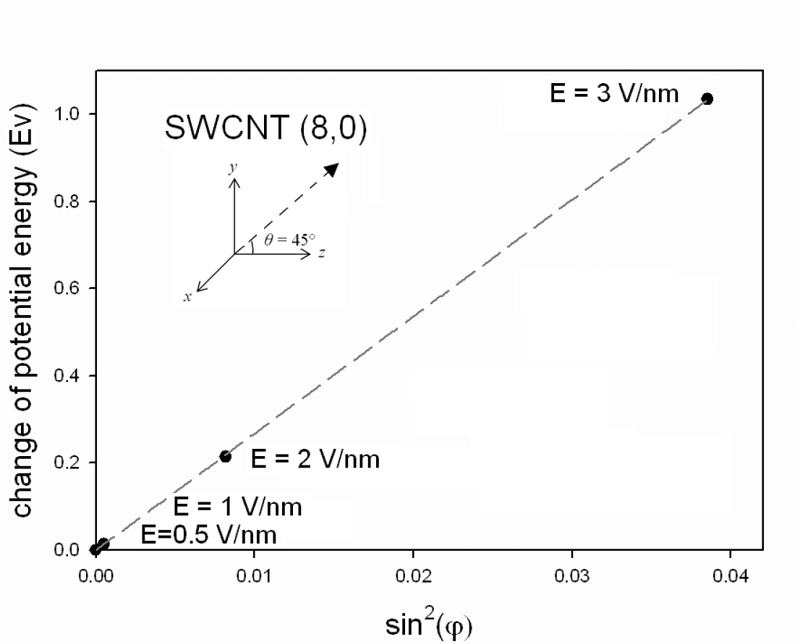}}
\caption{\label{fig:5}
(Color online) Change of potential energy $\Delta U^p$ of this (8,0) tube versus the square of $\sin(\varphi)$. 
The tube length is about $6.54$\,nm. $\theta=45^\circ$. The round points stand for the simulation data and the green dashed line represents the fitting line.}
\end{figure}

In Fig.\,\ref{fig:5}, we plot the change of the potential $U^{p}$ versus the square of $\sin(\varphi)$. It shows that $\Delta U^{p} \propto \sin^2(\varphi)$. Furthermore, we find also that the change of the global polarizability in the direction of the incident field $\bm{E}_0.\Delta \beta \bm{E}_0 / E^2$ is proportional to $\sin(\varphi)$ although the individual components of $\Delta \beta$ are not proportional to $\sin(\varphi)$ ($\Delta \beta_{yy} \propto \sin^2(\varphi)$ while $\Delta \beta_{zz} \propto -\sin^2(\varphi)$ and $\Delta \beta_{xx}$ remains almost constant). Combined with the fact that $\sin(\varphi)$ is almost proportional to the square of $E$ (Eq. \ref{eq:8}), this means that we also have $\Delta U^{elec} \propto \sin^2(\varphi)$, hence $\Delta U^{tot} \propto \sin^2(\varphi)$.

\begin{figure}[tb]
\centerline{\includegraphics[width=8cm]{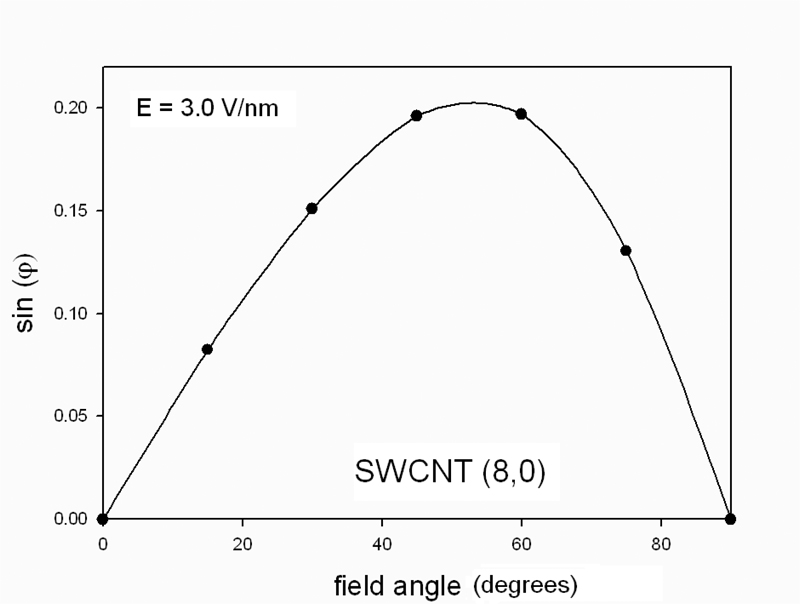}}
\caption{\label{fig:6}
$\sin(\varphi)$ versus electric field angles $\theta$ for a (8,0) tube. The tube length is about $6.54$\,nm. $E=3.0$\,V/nm. The round points stand for the simulation data.}
\end{figure}

In a second series of simulations, we studied the influence of the orientation of the external field on the response of CNTs. Fig.\,\ref{fig:6}  presents the relation between $\sin(\varphi)$ for a (8,0) SWCNT and the field angle $\theta$.

There is no obvious deflection effect found in CNTs when the applied electric field is either parallel or perpendicular to the main tube axis. In a global point of view, this is due to the fact that the induced molecular dipole of the tube is already aligned to the direction of the field in these two cases. That means the induced torque acting on the molecule is zero. The tube is therefore in equilibrium with the external field. We can also see in Fig.\,\ref{fig:6} that the deformation angle gets its maximum values when the field angle varies between $45$ and $60$ degrees. That means this tube can be most $'$efficiently$'$ bent in this field angle range. This range is biased towards 90 degrees because the axial polarizability of CNTs is always greater than the radial one.

Geometric effects are also examined. In several previous studies \cite{benedict-95}$^,$\cite{guo-04}$^,$\cite{kozinsky-06}, the static transverse polarizabilities of semiconducting CNTs are found to be proportional to the square of the tube diameter and almost independent of their chirality. We can therefore expect that the electrostatic deflections of two semiconducting SWCNTs of the same length and diameter but different chiralities are almost the same. In order to prove this expectation, we computed the induced deformation of a chiral (5,4) tube in a field with $\theta=45^\circ$ and $\textit{E} = 3.0$ V/nm. The tube length and radius are about $6.72$ nm and $0.31$ nm, respectively. The deformation angle $\varphi$ of this tube is about $11.85^\circ$. This value is close to that shown in Fig.\,\ref{fig:4} for a (8,0) tube in the same external field: $11.32^\circ$. The small difference between these two values may be attributed mainly to the fact that the elastic modulus of a zigzag SWCNT is slightly higher than that of a chiral or armchair one with same length and radius \cite{van-Lier-00}$^,$ \cite{zhaowang-06}. 

\begin{figure}[tb]
\centerline{\includegraphics[width=8cm]{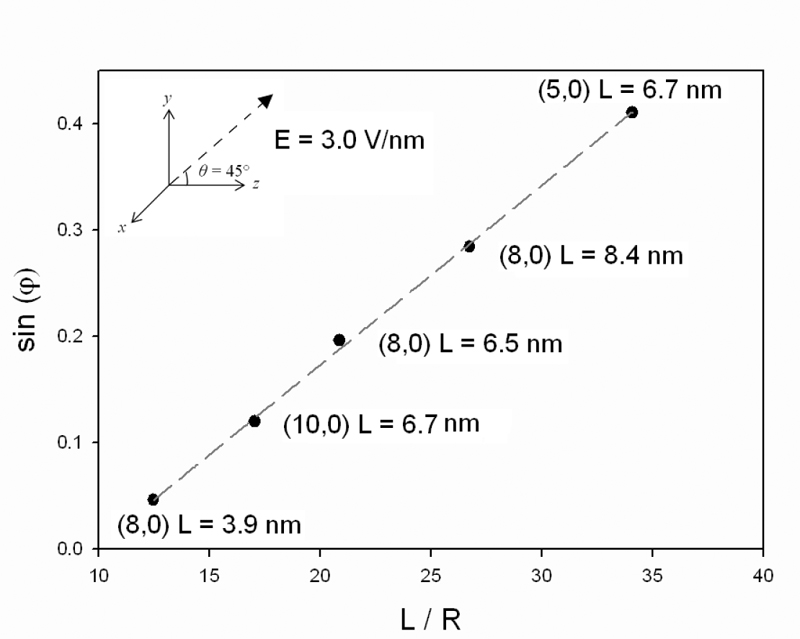}}
\caption{\label{fig:7}
(Color online) $\sin(\varphi)$ versus aspect ratio ($L/R$). $\theta=45^\circ$. $E=3.0$\,V/nm. The round points stand for the simulation data and the green dashed line represents the fitting line.}
\end{figure}

As shown in Fig.\,\ref{fig:7}, we found that $\sin(\varphi)$ is proportional to the aspect ratio of the tube defined as their length to radius $L/R$. For this external electric field, we obtain a relation almost linear between $\sin(\varphi)$ and the aspect ratio $L/R$ from our simulation results:

\begin{equation}
\label{eq:10}
\sin(\varphi)=B(L/R)+C
\end{equation}

where the values of $B$ and $C$ are about $0.0168 \pm 0.0003$ and $-0.163 \pm 0.008$, respectively. The mechanism of this phenomenon is rather complex: the deformation of a CNT depends both on the polarization and on its local mechanical resistance, the polarization depends on the external electric field and on the atomic polarizability, at the same time, the polarizabilities change following the deformation of the CNT. In our current work for building a global semi-phenomenological model for electrostatic deflection of CNTs, we are preparing some more simulations to try to get a better understanding on this point. 

In conclusion, we have demonstrated that a renormalized dipolar model coupled with the AIREBO potential can lead to realistic modelizations of the deflection of various nanotubes under homogeneous external electric fields. The sine of the deflection angle is proportional to the square of the external electric field, which is coherent with the conservation of the total energy of the system and the experimental data. This study also reveals an optimum deflection angle for an electric field making an angle between $45^\circ$ to $60^\circ$ with the original tube axis. This result is directly applicable in nanoelectronics where the nanotubes are oriented by means of electric fields in order to realize contacts between conductive plots. Furthermore, we also demonstrate the strong link existing between the deflection angle and the aspect ratio L/R of the tubes.

These results are interesting and are going to be studied in more details in forthcoming publications. For example, we have seen that the local electrostatic forces are strongly inhomogeneous along the tube but it seems that there exists a periodicity of its local variations that should be studied in more details for a wide variety of tubes. Furthermore, we would like to use a regularized monopole-dipole interaction model as developed by A. Mayer \textit{et al} \cite{mayer-06-01}, to be able to deal with metallic nanotubes containing free charges. Thus it could be possible to get a better understanding of the experimental results which suggest that metallic nanotubes can have a better response to electric fields than semiconducting ones. 

\begin{acknowledgments}
Prof. J.-M. Vigoureux and Dr. A. Mayer are gratefully thanked for fruitful discussions. This work was done as parts of the CNRS GDR-E 2756 and the NATO SFP project 981051. Z. W. acknowledges the support received from the region of Franche-Comt\'{e} (grant Nb 051129,91) and the Marie Curie conference and training courses grants.
\end{acknowledgments}

\end{document}